\documentclass[aps,pra,10pt,twocolumn,superscriptaddress]{revtex4}
\usepackage{amsmath}
\usepackage{amssymb}
\usepackage{bbm}
\usepackage{graphicx}
\usepackage{framed}
\usepackage{color}
\usepackage{hyperref}
\usepackage{epstopdf}
\usepackage{dsfont}
\usepackage{amssymb}
\usepackage{amsmath}
\usepackage{amsthm}
\usepackage{wrapfig}
\usepackage{relsize}
\usepackage{bm}
\usepackage[latin1]{inputenc}
\usepackage{enumitem}

\newlist{myitemize}{enumerate}{10}
\setlist[myitemize]{label*=\arabic*.,nosep,leftmargin=*}

\newcommand{\ba}{\begin{eqnarray}}
\newcommand{\ea}{\end{eqnarray}}
\newcommand{\pd}[2]{\frac{\partial #1}{\partial #2}}

\newcommand{\identity}{\mathlarger{
\mathds{1}}}

\begin{document}

\title{Optimal thermodynamic control in open quantum systems}

\author{Vasco Cavina}
\affiliation{NEST, Scuola Normale Superiore and Istituto Nanoscienze-CNR, I-56126 Pisa, Italy}
\author{Andrea Mari}
\affiliation{NEST, Scuola Normale Superiore and Istituto Nanoscienze-CNR, I-56126 Pisa, Italy}
\author{Alberto Carlini} 
\affiliation{Università degli Studi del Piemonte Orientale Amedeo Avogadro, Italy}
\author{Vittorio Giovannetti}
\affiliation{NEST, Scuola Normale Superiore and Istituto Nanoscienze-CNR, I-56126 Pisa, Italy}

\begin{abstract}

We apply advanced methods of control theory to open quantum systems and we determine finite-time processes which are optimal with respect to thermodynamic 
performances. General properties and necessary conditions characterizing optimal drivings are derived, obtaining bang-bang type solutions corresponding to control strategies switching between adiabatic and isothermal transformations. A direct application of these results is the maximization of the work produced by a generic quantum heat engine, where we show that the maximum power is directly linked to a particular conserved quantity naturally emerging from the control problem.
Finally we apply our general approach to the specific case of a two level system, which can be put in contact with two different baths at fixed temperatures, identifying the processes which minimize heat dissipation. Moreover, we explicitly solve the optimization problem for a cyclic two-level heat engine driven beyond the linear-response regime, determining the corresponding optimal cycle, the maximum power, and the efficiency at maximum power.
\end{abstract}

\maketitle 
\section{Introduction} Suppose that we have at disposal a quantum system that can be coupled with two heat baths of different temperatures: what is the most powerful heat engine that we can realize using the system as a working medium? 
For fixed initial and final states, what is the optimal finite-time transformation minimizing heat dissipation? These questions are at the basis of the current research activity in finite-time quantum thermodynamics \cite{Esposito2010c, Andresen2011, Anders2016}. Focusing on systems describable by quantum master equations, the aim of this work is to solve the previous maximization and minimization problems through the formalism of optimal control theory and, in particular, exploiting a quite useful technique  known as the {\it Pontryagin's minimum principle} (PMP) \cite{Pontryagin1962}.
 
Optimal control theory has proven to be a very powerful tool to face a wide range of problems in the quantum world,
from achieving fast and effectively adiabatic processes
with qubits or quantum oscilllators
\cite{Hoffmann2016, Carlini2006, Stefanatos2013, Stefanatos2015, Stefanatos2017, Deffner2014, Gueryodelin2014, Gueryodelin2015, Kosloff2017, Zhang2015}, to manipulating the relaxation time \cite{Tannor1999, Mukherjee2016, Martinez2016, Carlini2016} and the dissipation \cite{Rotskoff2016, Muratore2014, Sauer2013}
in open quantum systems.
The time-minimization problem turns out to be of great relevance also from a thermodynamic point of view, 
being fundamental in enhancing the performance of the quantum Otto engine \cite{DelCampo2014, Beau2016, Rezek2006}.
However, there are much more issues of quantum thermodynamics that can be tackled using optimal control,
from reaching the lowest achievable temperature and testing the third law of thermodynamics \cite{Rezek2006,Feldmann2010},
to the enhancement of finite-time thermal engines \cite{Bonanca2014,Feldmann1996,Esposito2010b,Esposito2010c}.
This last purpose has been sought in a variety of frameworks, e.g. in stochastic or harmonic oscillator quantum engines \cite{Schmiedl2007, Seifert2012, Brandner2015, Bauer2016, Abah2016},
in multi-level driven quantum systems \cite{Einax2014, Wang2012, Allahverdyan2010, Wang2013, Wu2014, Gelwaser2015}, and in systems with strong fluctuations \cite{Cavina2016}.
In the case of finite-time thermal engines, thermodynamics must be supported by a dynamical theory, that often consists of 
phenomenological equations \cite{Esposito2010a, Chen1994}, although more fundamental descriptions have been considered \cite{Cavina2017,Avron2012}.
In our analysis we focus on the paradigmatic  scenario of thermal engines which base their functioning on the possibility of modulating the interactions of a quantum system $S$ with 
a cold and a hot reservoir.  In this framework we show how the PMP can be used to identify optimal procedures  that allow for the minimization of the heat released or equivalently for the
maximization of the power produced into a cycle. 

\begin{figure}[t]
\includegraphics[width=0.4 \textwidth]{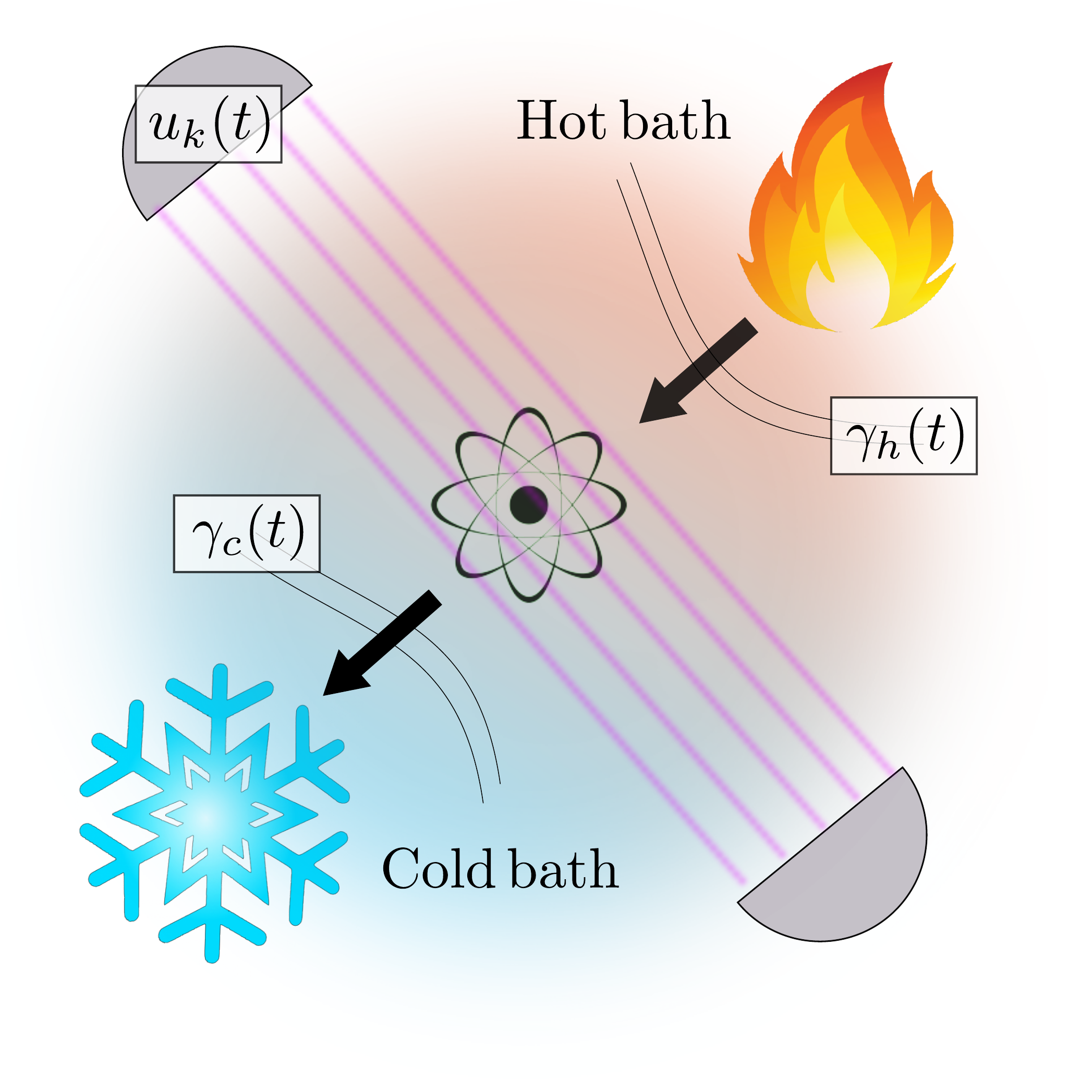}
\caption{Pictorial representation of the two-bath model: a system $S$ whose Hamiltonian $\hat{H}_{\bold{u}(t)}$ is driven via a collection of external control 
fields $\bold{u}(t)$,
evolves in time while being 
  coupled  with a cold bath of inverse temperature
$\beta_c$ and with a hot bath of inverse temperature $\beta_h$ through the couplings parameters $\gamma_c(t)$ and $\gamma_h(t)$ which can also be  externally controlled.}\label{fig:1}
\end{figure}
For this purpose we adopt the Markovian master equation approach \cite{Breuer2002} and describe the time evolution of the   density matrix $\hat{\rho}(t)$ of 
 $S$ in terms of 
a first order  differential equation  
\ba \label{dyn}  
\frac{d {\hat{\rho}}(t)}{dt} = \mathcal{L}_{\bold{u}(t)}[{\hat{\rho}}(t)] := - i [\hat{H}_{\bold{u}(t)}, \hat{\rho}(t)] + \mathcal{D}_{\bold{u}(t)}[{\hat{\rho}}(t)]\;, 
\ea
with  $\hat{H}_{\bold{u}(t)}$ and $ \mathcal{D}_{\bold{u}(t)}$ being, respectively, the (possibly time-dependent) system Hamiltonian  and the Gorini-Kossakowski-Sudarshan-Lindblad  dissipator~\cite{Lindblad1976, Breuer2002}   which gauges the thermalization process induced by  the reservoirs  connected to $S$. As implicitly indicated by the notation, 
 both these terms are assumed to exhibit a parametric dependence upon $t$,
 mediated via a collection of external control fields  represented by the real vectorial function $\bold{u}(t) := \{ u_1(t), u_2(t),\cdots\}$ which can be used to drive the system evolution e.g.  by changing its energy spectrum
 or by selectively switching on and off the couplings with the various thermal baths.
Within this setting 
our aim is to find the best  strategy that minimizes the mean heat $Q$ released by $S$ while evolving according to Eq.~(\ref{dyn}) for a fixed time interval  $[0,\tau]$, i.e.  the quantity
~\cite{Alicki1979, Kosloff2013, Anders2013}
\ba \label{heat} 
Q := - \int_0^{\tau} 
\Big\langle \hat{H}_{\bold{u}(t)} \; \mathcal{L}_{\bold{u}(t)}  [{\hat{\rho}}(t)] \Big\rangle dt \;, 
\ea
where here and in the following  we use  $\langle \cdots \rangle$ to represent the  trace operation. 
For cyclic processes this corresponds to maximizing the work performed by the system, i.e. the quantity
\ba \label{work}  
W :=-  \int_0^{\tau} 
\Big\langle   \hat{\rho}(t) \; \frac{d{\hat{H}}_{\bold{u}(t)}}{dt} \Big\rangle dt \;.
\ea
In the first part of this paper the problem will be considered in his entirety: we find a set of necessary minimum conditions that can 
be used to prove some interesting results, from the existence of a conserved quantity along the optimal trajectories to the derivation of a 
state equation for the exchanged work and heat.
Eventually the optimal solutions will be partially bang-bang
and we will provide a set of necessary conditions that have to be satisfied 
at the switching points. This behavior  is reminiscent of the familiar switching between isothermal and adiabatic transformations of a standard Carnot cycle, which is known to have an optimal efficiency in the quasi-static limit \cite{Huang2009}.
Finally we focus on the case of a two-level system which is sufficiently simple to be treated analytically  and, at the same time, of significant conceptual relevance. 
Indeed, thanks to the application of the PMP, we are able to give a complete characterization of the optimal cycle achieving the maximum power output.

\section{Minimization of released heat}
To find a stationary point of the functional (\ref{heat}) we
will apply the PMP \cite{Kirk2012} to the extended functional
\ba\label{extfunctional0}
{\cal{J}}&:=&Q + \int_0^{\tau} \;  \Big\{ \lambda(t) ( \langle\hat{\rho}(t) \rangle- 1)
\\ 
&&\qquad +\Big\langle \mathcal{\hat{\pi}}(t) \left ( \mathcal{L}_{\bold{u}(t)}[{\hat{\rho}}(t)]-\frac{d {\hat{\rho}}(t)}{dt} \right )\Big\rangle\Big\}\;dt  \nonumber \;.
\ea
In this expression  $\hat{\mathcal{\pi}}(t)$ is a self-adjoint operator of the same dimension of $\hat{\rho}(t)$, called \textit{costate}~\cite{NOTA}, and $\lambda(t)$ is a
scalar function, both acting as Lagrange multipliers that enforce, respectively, 
  the dynamical constraint (\ref{dyn}), and the normalization condition  $\langle \hat{\rho}(t)\rangle=1$.
Replacing~(\ref{heat}) into (\ref{extfunctional0}) we can conveniently express $\cal J$ as 
\ba\label{extfunctional}
{\cal{J}}=\int_0^{\tau} \; \Big\{ {\mathcal{H}}(t) -\Big\langle \hat{\mathcal{\pi}}(t)\frac{d {\hat{\rho}}(t)}{dt} \Big\rangle \Big\} \; dt\;.\label{lagr}
\ea
where
the function ${\mathcal{H}}(t)$  is  the  \textit{pseudo Hamiltonian} of the model defined as
\begin{eqnarray} 
{\mathcal{H}}(t):=\Big\langle(\hat{\mathcal{\pi}}(t) - \hat{H}_{\bold{u}(t)}) \mathcal{L}_{\bold{u}(t)}[\hat{\rho}(t)]\Big\rangle 
+ \lambda(t) ( \langle \hat{\rho}(t)\rangle - 1) \nonumber .\\ \label{HH} 
 \end{eqnarray} 
The PMP provides us with a set of necessary conditions that have to be satisfied by an optimal choice of the control parameters~\cite{Kirk2012} in order
to minimize $Q$. In particular 
 it implies that {\it i)} a non-zero costate $\hat{\pi}(t)$ exists such that
 \begin{eqnarray} 
 \frac{d{\hat{\rho}(t)}}{dt} = \label{H1} \pd{\mathcal{H}(t)}{{\hat{\pi}(t)}} \;,  \qquad 
 \frac{d{\hat{\pi}}(t)}{dt} = - \pd{\mathcal{H}(t)}{\hat{\rho}(t)}\;,
 \end{eqnarray} 
the  first  reducing to~(\ref{dyn}), the second  
describing instead  the time evolution of the costate (see Eq.\ \eqref{dyn2} of the Appendix). 
The PMP establishes also that 
 for all
 $t \in [0,\tau]$ the pseudo Hamiltonian $\mathcal{H}(t)$ 
{\it ii)} has to be  minimum with respect to the controls functions $\bold{u}(t)$, and {\it iii)} it  has 
to assume a constant value $\mathcal{K}$, i.e.
 \ba \label{cons} 
\mathcal{H}(t)= \mathcal{K}.
 \ea
Thanks to the above construction 
we can finally  express the minimum value of the heat released by the system in the following 
compact form 
\ba \label{heat1}  
Q_{min}  = \langle \hat{\pi}(0) \hat{\rho}(0)\rangle  - \langle \hat{\pi}(\tau) \hat{\rho}(\tau)\rangle - \int_0^{\tau} \lambda(t)dt,
\ea
with 
\ba \label{DEFLA}
\lambda(t) = - \Big\langle \hat{\rho}_{eq}(t) \frac{d{\hat{\pi}(t)}}{dt} \Big\rangle\;,
\ea
where now all the quantities on the right hand sides are computed on the optimal trajectories  fulfilling the PMP requirements and where
$\hat{\rho}_{eq}(t)$ is a fixed point of the super-operator  $\mathcal{L}_{\bold{u}(t)}$ (see the Appendix for details).

It is worth noticing that the above conditions hold true independently from the 
initial and 
final states $\hat{\rho}(0)$, $\hat{\rho}(\tau)$ which can be fixed later on~\cite{Kirk2012}. 
It is also still possible to both fix or leave the final time $\tau$ free 
and perform a further optimization on it.
A last remark is mandatory about the  regularity of the optimal trajectories.  
The control fields which provide the local  minima of the pseudo Hamiltonian~(\ref{HH}) need not to be differentiable, nor continuous 
(an irregular behaviour which is common in control theory and goes under the name of  bang-bang control~\cite{Stefanatos2013, Sussmann1982}).
For this reason, when solving Eq.~(\ref{H1})  for $\hat{\rho}(t)$ and $\hat{\pi}(t)$ we are forced to accept \textit{piecewise smooth} solutions 
divided by instantaneous switchings of the controls $\bold{u}(t)$, in which
the state and the costate have to be continuous, though a discontinuity in their first derivative is allowed (Weierstrass-Erdmann corner conditions~\cite{Kirk2012}).

\section{Maximum power and physical meaning of $\mathcal K$} 
The quantity $\mathcal K$ appearing in Eq.~(\ref{cons})  corresponds to a constant of motion (the analogue of energy in Hamiltonian mechanics) but,  apart from providing a convenient parametrization of the optimal solutions, its physical interpretation might appear quite obscure.  
The situation changes however 
if, instead of minimizing the dissipated heat, we minimize its corresponding  emission rate 
$R := {Q}/{\tau}$ by optimizing also over $\tau$.
Indeed recalling Eq.~(\ref{lagr}) and the general law for the variation of an ``action" functional with respect to $\delta \tau$ (see, e.g., Ref.  \cite{Kirk2012}) we obtain that, on-shell (i.e. when Eq.~(\ref{dyn}) and $\langle\hat{\rho}(t)\rangle=1$ are satisfied) and for fixed initial and final states, the following holds:
$\delta {\mathcal{J}}=\delta Q = \mathcal{H} (\tau) \delta \tau = \mathcal{K} \delta \tau$.
Accordingly, the  variation of $R$ can be expressed as  
\begin{equation}\label{rmin}
\delta R_{\rm min}= \frac{\delta Q}{\tau}- \frac{Q \delta \tau }{\tau^2}= \frac{\delta Q}{\tau}- \frac{R_{\rm min} \delta \tau }{\tau}=(\mathcal K-R_{\rm min})\frac{\delta \tau}{\tau},
\end{equation}
which nullifies if and only if $R_{\min}=\mathcal K$. It is worth stressing that the above analysis does not tell us the explicit value of $R_{min}$: it is just a formal identity which only shows that the latter coincides with the value of  ${\cal K}$ associated with the trajectory that yields the minimal rate. Yet Eq. (\ref{rmin}) allows to establish the following lower bound 
\begin{equation}\label{maxPower}
R_{\min} := \min_{\substack{{\rm all \; protocols},  \\  {\rm all\ } \tau>0 }} \frac{Q}{\tau}\ge {\cal K}^*:=  \min_{ \mathcal K \in A }   \mathcal K,
\end{equation}
where $A$ denotes the accessible region for the parameter $\mathcal{K}$ (i.e. the set which contains the values of ${\cal K}$ that allow for an integration of the equations of motion (\ref{H1}) which are consistent with the PMP constraints {\it ii)} and {\it iii)} and with the assigned initial and final conditions of the problem).
Equation~(\ref{maxPower}) is remarkable since it allows to replace the minimization of the functional $R$ with respect to all possible protocols, with a much simpler, single-parameter minimization.
 Notice also that for cyclic processes we have $W=-Q$, and so the maximum power achievable from the quantum system $S$ when exploited as a thermal engine is bounded as:
\begin{eqnarray}
P_{\max} := \max_{\substack{{\rm all \; protocols},  \\  {\rm all\ } \tau>0 }} \frac{W}{\tau}\le -{\cal K}^*\;. \label{MAX} 
\end{eqnarray} 

Moreover, if the critical solution with $\mathcal K=\mathcal K^*$ is such that the corresponding rate is equal to $\mathcal K$, then this is necessarily the optimal solution and the previous bounds \eqref{maxPower} and \eqref{MAX} are saturated. 
As we are going to show later, this is exactly what happens for the optimization of a two-level system heat engine. For general systems, we expect a saturation of both the inequalities \eqref{maxPower} and \eqref{MAX} whenever there exists a sufficiently regular infinitesimal cycle with conserved quantity $\cal{K}=\cal{K}^*$.  Indeed, in the limit $\tau \rightarrow 0$, the heat rate along critical trajectories can be expressed, using Eq. (\ref{extfunctional}), as
\begin{align} \label{Rshort}
R=& \frac{{\cal{J}}}{\tau}= \lim_{\tau \rightarrow 0} \frac{1}{\tau}\int_0^{\tau} \; \Big\{ {\mathcal{H}}(t) -\Big\langle \hat{\mathcal{\pi}}(t)\frac{d {\hat{\rho}}(t)}{dt} \Big\rangle \Big\} \; dt\;  \nonumber  \\ 
   = &\: \mathcal{H}(\tau) - \hat{\mathcal{\pi}}(0)[ \rho(\tau)-\rho(0)]/\tau= \mathcal{H}(\tau)=\mathcal K,
\end{align}
where we used that in cyclic process $\rho(0)=\rho(\tau)$ and ${\hat\pi}(t) \simeq \mathrm{const}$.

To derive the previous results we have implicity assumed that the energy cost of switching on/off the system-bath interactions is negligible, which is a standard approximation in the weak coupling limit.
Nevertheless, in an infinitesimal cycle the number of switchings goes to infinity and their contribution to the work may be finite, vanishing only for specific forms of the interaction Hamiltonian ({\it e.g.} a partial swap \cite{Scarani2002}).
A more accurate discussion would involve a microscopic characterization of the system-bath interactions \cite{Newman2017, Perarnau2018} which goes beyond the purposes of the present work.

\section{Two baths scenario} 
We now focus on  the paradigmatic case where $S$ is directly coupled to only two baths, a cold reservoir of inverse temperature $\beta_c$, and
a hot reservoir of inverse temperature $\beta_h$, see Fig. 1. Accordingly we take 
\ba \label{dyn1}  
\mathcal{D}_{\bold{u}(t)}[\cdots]=   \gamma_c(t) \mathcal{D}_{\bold{u}(t)}^{(c)}[\cdots]  + \gamma_h(t) \mathcal{D}_{\bold{u}(t)}^{(h)} [\cdots]\;,
\ea
with  $\mathcal{D}_{\bold{u}(t)}^{(c)}$ and $\mathcal{D}_{\bold{u}(t)}^{(h)}$ being the dissipators 
describing the thermalization processes induced by the two reservoirs
and with  $\gamma_{c,h}(t)$ being  the corresponding  damping  rates which we consider as dedicated elements of the control fields set. 
The complete-positivity of the dynamics implies that these two damping parameters are non-negative. 
On top of this, while keeping unconstrained the  remaining control fields, we 
restrict ourselves to the case in which the total damping rate is equal to a given positive constant $\Gamma$, i.e. $\gamma_c(t) + \gamma_h(t) = \Gamma$. 
This condition  enforces a physically motivated termalization time scale, preventing the emergence of trivial solutions, and is ideally suited to the case of typical thermal machines working between two different temperatures.
As a matter of fact under this assumption the minimization of Eq.~(\ref{HH}) with respect to   $\gamma_{c,h}(t)$ can be easily performed, yielding only two extremal control strategies in which the system is selectively coupled at maximum rate $\Gamma$ with either the cold or the hot thermal bath.  These two possible regimes can be activated depending upon the sign of the functional
\begin{eqnarray} 
{\cal A}_{\bold{u}(t)}(\hat{\pi}(t), \hat{\rho}(t)) \!:= \Big\langle\!\!(\hat{\pi}(t) -\hat{H}_{\bold{u}(t)})(\mathcal{D}_{\bold{u}(t)}^{(h)}-\mathcal{D}_{\bold{u}(t)}^{(c)}) [\hat{\rho}(t)]\!\Big\rangle.
\nonumber \\ \label{H3i} 
\end{eqnarray} 
In particular the choice $\gamma_c(t) = \Gamma, \: \gamma_h(t) = 0$, corresponding to the case where $S$ only interacts with the cold bath, is available whenever  ${\cal A}_{\bold{u}(t)}  \geq  0$  (cold isotherms \cite{NOTA2}), while  the choice $\gamma_c(t) = 0,  \: \gamma_h(t) = \Gamma$  is available if 
${\cal A}_{\bold{u}(t)}  \leq 0$  (hot isotherms). 
Following a standard bang-bang approach, the optimal trajectory can  then be obtained by  dividing the total time interval  $[0,\tau]$ into 
an ordered  sequence of intermediate steps 
where one of the above behaviours applies -- the explicit values of  the controls ${\bold{u}(t)}$  being fixed by 
solving the corresponding  Eq.~(\ref{H1}) under the  
PMP conditions {\it ii)} and {\it iii)}. 
These isothermal evolutions are separated 
by intermediate switching times 
where, with $\hat{\rho}(t)$ and $\hat{\pi}(t)$  still preserving continuity,  $S$ may experience {\it adiabatic jumps} on  $\bold{u}(t)$ 
which effectively result in an instantaneous decoupling of the system  from the thermal baths (adiabats).      
The optimal switching times can be found by solving the set of algebraic equations derived from the continuity of the state/costate variables.
From a numerical point of view this is an easy task to accomplish, at least if compared to an {\it ab initio} numerical optimization 
over the whole family of bang-bang trajectories.

\section{The two-level system case} 
As an example of a two bath model for which the PMP optimization can be explicitly solved  we consider the case in which $S$ is a two level system
driven by a time dependent Hamiltonian that has constant eigenvectors  $\{ |0\rangle, |1\rangle\}$, but an energy gap  $u(t)\geq 0$
 which can  be externally modulated, 
i.e. 
$\hat{H}_{u(t)} := u(t)|1\rangle \langle 1|$.
For dissipators we take the super-operator defined by the mapping
\begin{equation} \label{master}  
\mathcal{D}_{{u}(t)}^{({c,h})}[ \hat{\rho}(t)]  = \hat{\eta}_{\beta_{c,h}}(t) - \hat{\rho}(t), 
\end{equation}
where $\hat{\eta}_{\beta}(t):= [e^{-\beta u(t)}|1\rangle\langle 1| + |0\rangle\langle 0|]/(e^{-\beta u(t)} +1)$
 is the instantaneous Gibbs state of $\hat{H}_{u(t)}$ with inverse temperature $\beta$ (for $\beta>0$). 
This model is similar to one considered by Esposito et al. in \cite{Esposito2010b,Esposito2010c}, for which our approach  allows now a systematic and rigorous  solution of the optimal trajectory problem.
The restriction to a non-rotating Hamiltonian is a necessary assumption for minimizing heat dissipation, at least when the dynamics is 
induced by a dissipator of the form (\ref{master}).
This can be proven by constructing a set of equations {\it ad hoc} for the coherent case, starting from the PMP conditions 
introduced in the first section of the actual work.
However, this discussion involves many technical issues and we delay it 
to a next publication.

A physical implementation of Eq.~(\ref{master}) is realizable {\it e.g.}\ with a single level quantum dot in contact with
two fermionic heat baths, in the wide band approximation \cite{Esposito2010b, Harbola2006}.
In this case the low energy level $|0\rangle$ is associated with an empty dot, while the excited level  $|1\rangle$ is associated with a dot populated by one fermion.
For the sake of simplicity, as in Ref.~\cite{Esposito2010b}, we 
 shall focus on the case where the initial state of $S$ exhibits no coherences in the energy eigenbasis. Under these conditions Eq.~(\ref{dyn}) 
ensures that the density matrix of the system maintains  a diagonal form at all $t$, allowing us to express it as  
 $\hat{\rho}(t) : = {p}(t) |1\rangle\langle 1| + (1-{p}(t)) |0\rangle\langle 0|$ where ${p}(t)$ is  the probability for $S$ of being in the excited level.  Replacing this into (\ref{extfunctional}) and exploiting the  fact that we have the gauge freedom to assume  $\hat{\pi}(t)$ to be traceless~\cite{NOTA} it then  follows 
 that we can also neglect the coherence contributions to the costate, writing it as $\hat{\pi}(t)= q(t)( |0\rangle\langle 0| - |1\rangle\langle 1|)$, with $q(t)$ to be determined by solving the
 dynamical equation~(\ref{H1}). 
\begin{figure*}[t!]
\includegraphics[width=0.7 \textwidth]{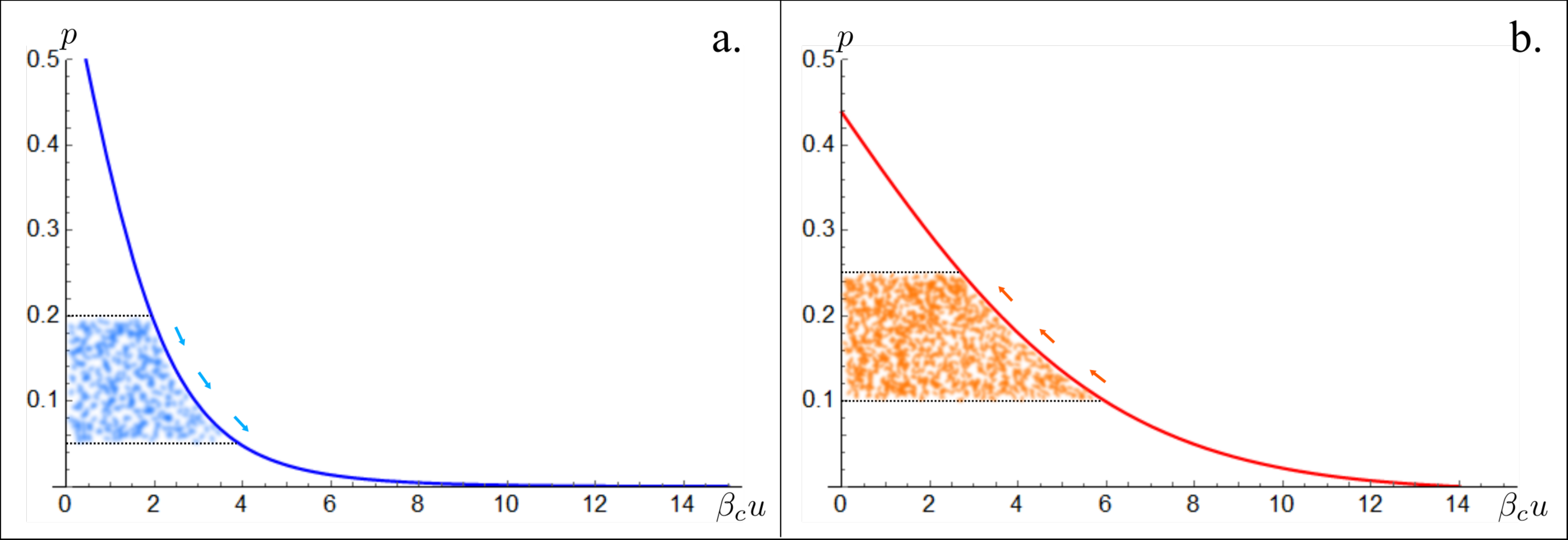}
 \caption{{\bf a.} Form of the cold (blue) isotherm when $\beta_h=0.3 \,\beta_c$ and $\mathcal{K}=-0.05\, \Gamma/\beta_c$.
{\bf  b.}  Form of the hot (red) isotherm for the same set of parameters.
 The area below the function $p(\beta_c u)$ is equal to the heat released (2a, colored in light blue) or the heat absorbed ({2b}, colored in orange) for the cold and hot isotherm respectively, measured in units of $\beta_c^{-1}$.
 The arrows show the direction of the dynamics.}\label{fig:2A}
 
\includegraphics[width=0.7 \textwidth]{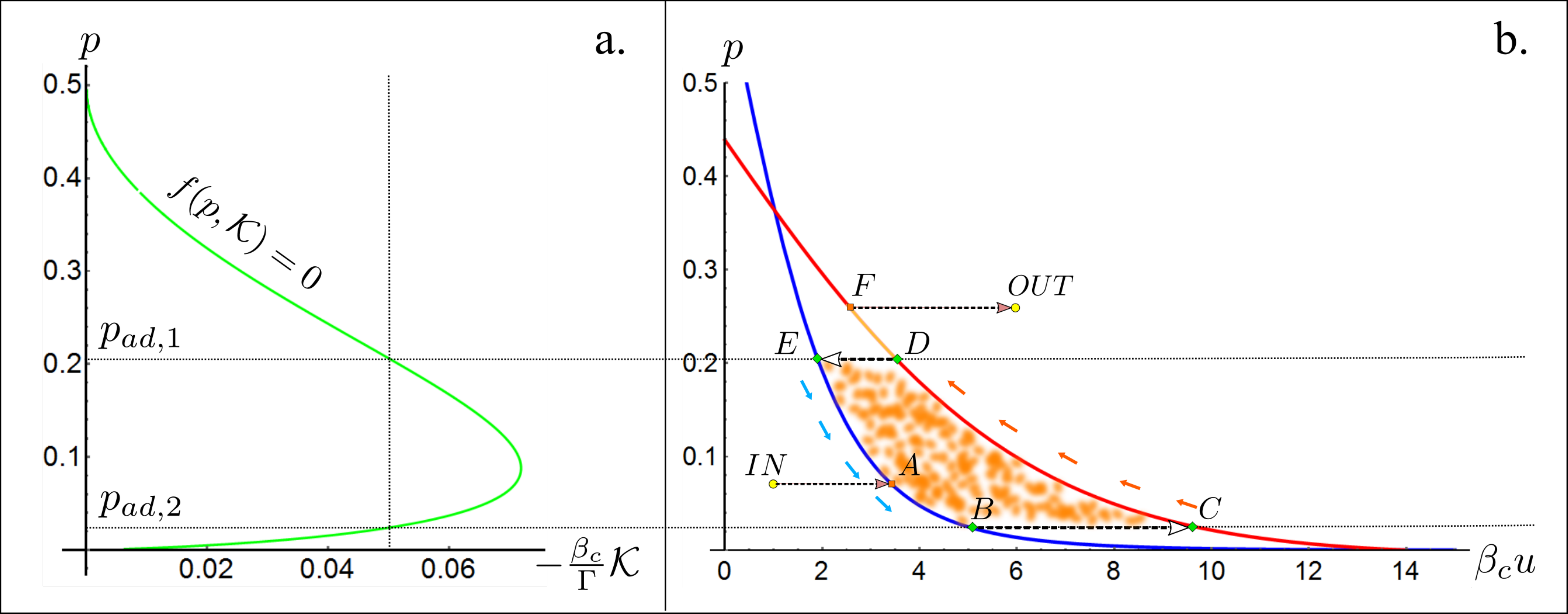}
\caption{{\bf a.} Profile of the zero-contour of $f({p},\mathcal{K})$ when $\beta_h=0.3\,\beta_c$ determining the condition for adiabatic jumps. For ease of representation we plot the contour of $f$ as a function of $-\mathcal{K}$ and we omit the
dependence of $p_{ad, 1}(\mathcal{K})$ and $p_{ad, 2}(\mathcal{K})$ by $\mathcal{K}$.
{\bf b.} General form of the optimal trajectories for $\beta_h=0.3\,\beta_c$ and  $\mathcal{K}=-0.05\, \Gamma/\beta_c$.
Curved blue (red) lines are cold (hot) isotherms.
The two horizontal lines correspond to the values $p_{ad, 1}$ and $p_{ad, 2}$ when the system undergoes intermediate adiabatic jumps (represented with horizontal black arrows).
The area enclosed by the cycle is colored in orange and is equal to the heat absorbed during the cycle.} \label{fig:2B}
\end{figure*}

Under these assumptions it is possible to explicitly solve the optimal trajectory problem. 
For this sake we recall that our solutions can be 
parametrized by the constant of motion $\mathcal{K}$, defined in Eq.~(\ref{cons}), 
that turns out to be always negative for the model in consideration (see the Appendix).
Specifically, introducing the adimensional quantities $\mu_{c} := -\sqrt{-(\beta_{c} \mathcal{K})/\Gamma}$ and  $\mu_{h} := 
   \sqrt{-(\beta_{h} \mathcal{K})/\Gamma}$,
   it follows that the evolution of the density matrix of $S$ 
   along an optimal cold (resp. hot)  isothermal step is given by 
  \begin{eqnarray} 
\label{sol1} 
 {p}(t) &=& \frac{1- \mu_{c,h} \; x_{c,h}(t)}{1+x_{c,h}^2(t)}. 
\end{eqnarray}
The dependence upon the control parameter $u(t)$  in (\ref{sol1}) is gauged through the function $x_{c,h}(t) :=  \exp \left [\beta_{c,h} u(t)/2 \right ]$. This  is determined by the differential equation 
 \ba  \label{tdiff} 
\frac{d {x}_{c,h}(t)}{dt}  = \frac{ \Gamma[x_{c,h}^2(t)+1]\; x_{c,h}(t)}{\left [x_{c,h}^2(t) - 2\frac{x_{c,h}(t)}{\mu_{c,h}} -1\right ]},
  \ea
that can be integrated for assigned boundary conditions (see the Appendix).
In the remainder of this paragraph we will discuss a specific example in which, for the sake of simplicity, $\Gamma^{-1}$ and $\beta_c^{-1}$ are fixed and adopted as units of time and energy respectively,
such that the rate $\mathcal{K}$ is measured in units of $\Gamma/\beta_c$. 
Examples of the isothermal trajectories (\ref{sol1})  in the plane $({p},\beta_c u)$ are presented in Fig.~\ref{fig:2A}  for $\beta_h=0.3\;\beta_c$ and $\mathcal{K}=-0.05\;\Gamma/\beta_c$.
 In the picture it is also reported the associated amount of the heat  released by $S$:  for the model we are considering, this quantity  admits an explicit, yet cumbersome, analytic expression and a simple geometrical interpretation as minus the area subtended by the curve $u(p)$ between the initial and final 
 points of the corresponding trajectory  (see the Appendix). 
 The general optimal process can involve also adiabatic jumps of the field $u(t)$ in which $S$ moves from one isothermal trajectory (say cold) to the other (say hot).
 In such a process the continuity of $p(t)$ and $\pi(t)$ has to be preserved, and
we formulate this requirement introducing a function $f(p,\mathcal{K})$ whose zero contour associates each value of 
 ${\cal K}$ to the values of $p$ in which an adiabatic quench is allowed (cfr. Eqs. (\ref{ad1}-\ref{adiabfin}) of the Appendix).\\
\phantom{xx}An example for the case with $\beta_h=0.3\,\beta_c$ is provided in Fig. \ref{fig:2B}a:
the model admits a threshold value ${\cal K}^*$ for the parameter ${\cal K}$, that is exactly the quantity introduced in Eq.~(\ref{maxPower}).
For values of ${\cal K} < {\cal K}^*$  no adiabatic jumps are allowed
in the construction of the optimal trajectory  minimizing the global heat released in the process. 
On the contrary for  values of ${\cal K}\geq {\cal K}^*$, such jumps  
may occur whenever along an isothermal trajectory the probability $p(t)$ assumes two specific values $p_{ad, 1}({\cal K})$, 
$p_{ad, 2}({\cal K})$ corresponding to the zeros of the function $f(p,{\cal K})$ defined in Eq.~(\ref{adiabfin}).
Fig. \ref{fig:2B}b shows the form of the optimal trajectories for the values of the parameters specified above and $\mathcal{K}=-0.05\,\Gamma/\beta_c$.
Let us consider, for instance, an initial condition with $u(0)=\beta_c^{-1}$ and ${p}(0)=0.07$ (point $IN$ in Fig.  \ref{fig:2B}b)
and suppose that we want to reach the final configuration with $u(\tau)= 6 \, \beta_c^{-1}$ and $ p=0.26$ (point $OUT$ in Fig. \ref{fig:2B}b) following an optimal trajectory.
We first note that, since the direction of the dynamics in the cold/hot isotherms is respectively fixed to be downwards/upwards (see Figs.  \ref{fig:2A}), it can be shifted 
only through an adiabatic quench.
Moreover, as we have already noted, for any fixed constant of the motion $\mathcal K$, only two such optimal adiabats are allowed, i.e. only the pieces of trajectories $B$-$C$ and $D$-$E$ in Fig. \ref{fig:2B}b. 
With all of this in mind, the shortest path to reach the final configuration turns out to be $IN$-$A$-$B$-$C$-$F$-$OUT$ in Fig. \ref{fig:2B}b. 
However, we can also construct other optimal trajectories if, once we get to the point $D$, we choose to perform a cycle $D$-$E$-$B$-$C$ before continuing towards the point $F$.
We can add as many cycles as we want, the price to pay will be an increment of the final time $\tau$.
Accordingly we can decompose the total duration of the protocol and the corresponding heat released by $S$ as 
\ba \label{dect} 
\tau &=& \tau_c(\mathcal{K}) + \tau_h(\mathcal{K}) + N \tau_{cycle}(\mathcal{K}),  \\
\label{decQ}  
Q &=& {Q}_c(\mathcal{K}) +  {Q}_h(\mathcal{K}) + N {Q}_{cycle}(\mathcal{K}),  
\ea
where $\tau_{cycle}(\cal{K})$ and ${Q}_{cycle}(\mathcal{K})$ are the contributions associated with a complete inner cycle and 
$\tau_c(\mathcal{K}), \tau_h(\mathcal{K})$
and  ${Q}_c(\mathcal{K})$, ${Q}_h(\mathcal{K})$ those associated with the part of the trajectories which connect the inner cycle to the initial and final conditions
 (all quantities admitting explicit analytical expressions). It turns out that, for fixed $N$, the terms on 
 the r.h.s. of Eq.~(\ref{dect}) are increasing functions of ${\cal K}$, while the terms on
 the r.h.s. of Eq.~(\ref{decQ})  are decreasing functions of  ${\cal K}$ (see the Appendix). 
 Accordingly one would be tempted to minimize $Q$ by taking the larger value of the parameter ${\cal K}$. 
 Yet, as we reduce ${\cal K}$,
the terms on the r.h.s. of Eq.~(\ref{dect}) decrease to the point that, for fixed $\tau$, it is possible to 
increase $N$ by one, allowing the start of a new 
inner cycle. When this happens  the heat in Eq.~(\ref{decQ}) acquires an extra negative correction resulting in a net decrease of the total released heat $Q$: 
choosing lower values of ${\cal K}$ appears hence to be the optimal choice. This is particularly evident when we consider the asymptotic limit $\tau\rightarrow \infty$.
In this situation the 
initial and final probabilities ${p}(0)$ and ${p}(\tau)$ do not play any role, as the energy contributions ``outside of the cycle"
in Eq.~(\ref{decQ}) become negligible due to the divergency of $N$.

Since the optimal protocol collapses to an infinite succession of identical cycles, the maximization of the total power output (minimization of the total heat emission rate) finally reduces to searching for the cycle in which  $P$ ($R$) assumes the highest (lowest) possible value.
As we know from the previous discussion about the physical link between $\mathcal K$ and the maximum power, a good candidate for the potentially optimal cycle is the one for which $\mathcal K$ assumes its minimum accessible value $\mathcal{K}^*$. As we are going to show in the next section, such cycle saturates both inequalities \eqref{maxPower} and \eqref{MAX}, implying that it achieves the maximum power.

\section{Maximum power heat engine for a two level system} 
Setting $\mathcal{K}= \mathcal{K^*}$, the two adiabatic switching points 
collapse to the same critical value $p^*:=p_{ad, 1}({\cal K}^*)=p_{ad, 2}({\cal K}^*)$.
\begin{figure*}[t!]
\includegraphics[width=0.7 \textwidth]{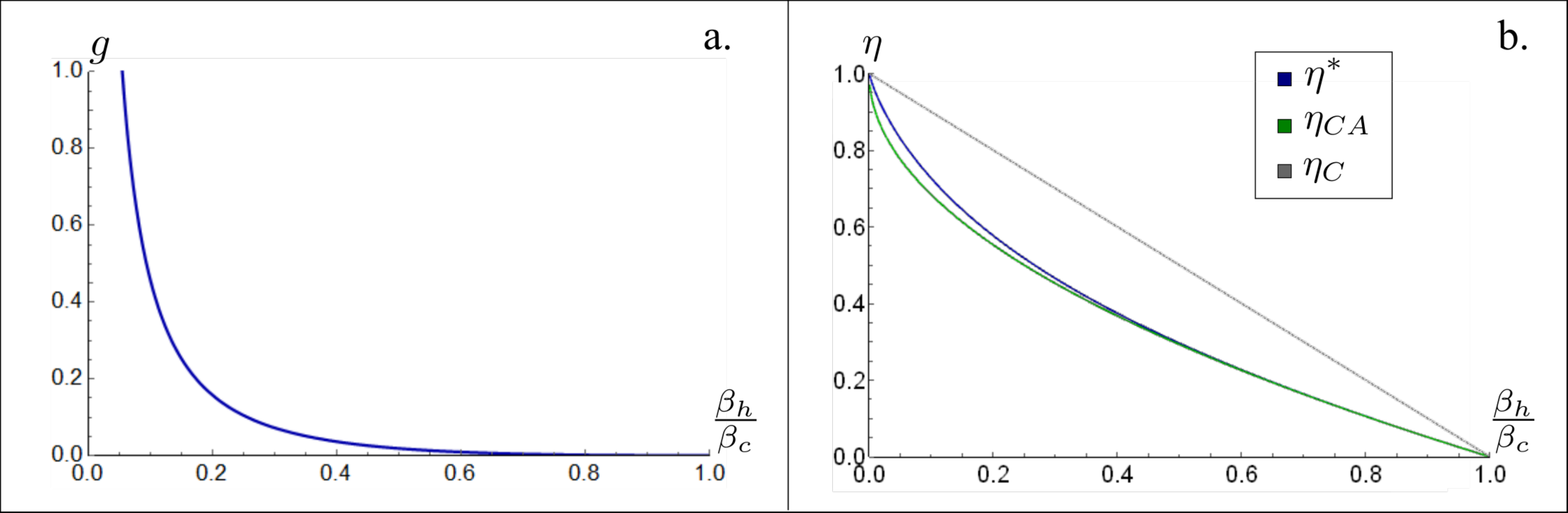}
 \caption{{\bf a.} Dimensionless function $g$ proportional to the minimum heat dissipation rate according to $\mathcal{K}^*= - (\Gamma/\beta_c)g( \beta_h/\beta_c)$.
 For cyclic processes, $g( \beta_h/\beta_c)$ corresponds to the maximum power (in units of $\Gamma/\beta_c$) achievable by the two-level system heat engine. 
  {\bf b.} Efficiency at maximum power $\eta^*$ derived in Eq.\ \eqref{effic} compared with the Carnot efficiency $\eta_C$ and the Curzon-Ahlborn efficiency $\eta_{CA}$.} 
\label{fig:3}
\end{figure*}
This yields
an asymptotically vanishing amount of the corresponding heat released per cycle ($\lim_{{\cal K} \rightarrow {\cal K}^*} {Q}_{cycle}(\mathcal{K})=0$) 
and makes the heat rate an indeterminate form (a similar phenomenon has been observed in the weak dissipation limit \cite{Esposito2010c}). Physically this optimal regime corresponds to an infinitesimal cycle 
in which the quantum state is almost unaffected but, at the same time, the Hamiltonian is subject 
to finite quenches where $u$ oscillates back and forth between two extremal values $u^*_{c,h}$ associated with the cold/hot (isochoric) processes (see the Appendix), as in a Otto cycle. The corresponding optimal time evolution of the cycle and of the control function $u(t)$ are plotted in Fig. \ref{fig:4}c.
What is more, since the cycle is also infinitesimal with respect to its time duration, from equation \eqref{Rshort} we also have $R=\cal{K}^*$ and both bounds  \eqref{maxPower} and \eqref{MAX} are saturated. This means that
the infinitesimal Otto cycle is optimal and the corresponding maximum power is given by $P_{\max}=-\cal{K}^*$.

As discussed in the Appendix, the minimum rate $\mathcal{K}^*$ depends upon the bath temperatures and the maximum coupling as in
\begin{eqnarray} \label{kstar}
{\cal K}^* = -\frac{\Gamma}{\beta_c} g \left (\frac{\beta_h}{\beta_c} \right )\;,
\end{eqnarray}
with $g(z)$ a dimensionless function that measures the power output of the process and has been computed numerically in Fig. \ref{fig:3}a.
For $z\rightarrow 1$ this quantity, and hence ${\cal K}^*$, nullifies:
this is the linear response regime where the two reservoirs have similar temperatures and where the maximum power (\ref{MAX}) asymptotically vanishes (see {\it e.g.} \cite{Esposito2010c}).
In the opposite limit  $z\rightarrow 0$ (which also includes the zero temperature limit for the cold bath), the function $g(z)$  is maximized and behaves as $g(z)\approx \theta/z$ , with $\theta\approx 0.06961$ being a dimensionless constant, yielding 
\begin{equation}
\label{PPP} P_{\rm max}\approx 0.06961\;  \frac{\Gamma} {\beta_h}\;,
\end{equation}
corresponding to the ultimate upper limit for the power achievable by a two-level system heat engine working on a fixed temperature gradient.
The analytic treatment of the limit $z \rightarrow 0$ is provided in the Appendix.
Finally from the definition \eqref{heat} we can also compute the infinitesimal heat dissipated in the cold and hot isotherms of the optimal cycle  
obtaining  $dQ_c= u_c^* d{p}\ge 0$, 
$dQ_h= -u_h^* d{p}\le 0$.
The efficiency at maximum power has therefore the following simple expression
\ba  \label{effic}
\eta^* = 1- \frac{u_c^*}{u_h^*},
\ea
which remarkably corresponds to the same efficiency of a standard Otto cycle subject to complete thermalizations.
Equation~(\ref{effic}) has been computed after solving numerically the maximum power equations (see the Appendix) and is plotted in Fig. \ref{fig:3}b along with the Carnot efficiency $\eta_C=1-\beta_h/\beta_c$ and the Curzon-Ahlborn efficiency $\eta_{CA}=1-\sqrt{\beta_h/\beta_c}$ \cite{Curzon1975}. 
We observe that the exact result is well approximated by the Curzon-Ahlborn efficiency which is known to be a universal feature common to many classical and quantum heat engines \cite{Andresen2011,Esposito2010a}. The slight discrepancy between $\eta^*$ and $\eta_{CA}$ can be ascribed to the fact that our solution is exact while the Curzon-Ahlborn efficiency is usually derived under different approximations, {\it e.g.}\ in the linear-response regime \cite{Benenti2017} or in the slow-driving limit \cite{Cavina2017}. By direct inspection of Figs.\ \ref{fig:3}a and \ref{fig:3}b, we also observe that $\eta^*$ converges to $\eta_{CA}$ in the weak dissipation limit $g\rightarrow 0$, consistently with the results of \cite{Esposito2010c}.
\section{Conclusions}
In this work we presented a  general theory of optimal control specifically designed for optimizing thermodynamic cost functions. Exploiting Lagrangian methods and Pontryagin's minimum principle we systematically derived in a general way a set of necessary analytical conditions characterizing thermodynamic processes with minimum heat dissipation. We found that optimal solutions can be parameterized by a particular conserved quantity $\mathcal K$  which is strongly related to the minimum heat dissipation rate and, for cyclic processes, to the maximum work power. 
We also proved that the controls in the optimal dynamics are of bang-bang type switching between isothermal and adiabatic evolutions.
 We applied the formalism to the paradigmatic case of a two-level system whose dynamics is governed by a thermalizing master equation, modeling the coupling with two different heat baths. For arbitrary initial and final conditions, we determined the class of critical thermodynamic processes minimizing heat dissipation.  When the system is used as a cyclic heat engine, we explicitly found that the maximum power is obtained for an infinitesimal Otto-like cycle performed around a particular non-equilibrium quantum state.
For what concerns a potential physical implementation of our model good candidates are, for instance,  single level quantum dots \cite{Ledentsov1998,Esposito2010b}.
Indeed their stability and the high tunability may allow to realize the optimal processes studied in this work directly in the laboratory.
From a technical point of view we have highlighted the versatility of the PMP approach to optimal control of quantum thermodynamics, paving the way to further applications in the field, e.g. the minimization of entropy production in open quantum systems \cite{Breuer2002}.
The entropy production functional is strictly connected to the heat functional (\ref{heat}), to which is substantially equivalent in the single bath scenario, 
although its non linear dependence on the state could lead to non trivial deviations from the 
results of the present paper when considering more thermal baths.
\\

\appendix
\section{The Pontryagin's minimum principle}
By differentiating the pseudo-Hamiltonian with respect to $\hat{\rho}(t)$ one can easily verify that the second of the PMP equations~(\ref{H1}) 
corresponds to 
 \begin{eqnarray}   \label{dyn2} 
\frac{d {\hat{\pi}}(t)}{dt}  = - \left \{{\mathcal{L}}_{\bold{u}(t)}^{\dag}[ \hat{\pi}(t)- \hat{H}_{\bold{u}(t)} ]
 +  \lambda(t) \identity \right \} \;, 
 \end{eqnarray} 
 with  ${\mathcal{L}}_{\bold{u}(t)}^{\dag}$ being the adjoint of the generator ${\mathcal{L}}_{\bold{u}(t)}$. 
By multiplying both sides of (\ref{dyn2})
by a fixed point state $\hat{\rho}_{eq}(t)$ of the super-operator  $\mathcal{L}_{\bold{u}(t)}$ and then taking the trace,  it immediately
yields Eq.~(\ref{DEFLA}). 
Equation~(\ref{heat1}) instead can be obtained by using (\ref{dyn}) and (\ref{dyn2}) to express the time derivative of
$\langle \hat{\pi}(t) \hat{\rho}(t)\rangle$ and integrating it over the time interval $[0,\tau]$. 
Notice also that the PMP requirement {\it ii)} can be translated into a set of necessary conditions, 
by  imposing stationarity of ${\cal H}(t)$ with respect to variation of   the control fields, i.e. 
 \begin{eqnarray}  
\Big\langle (\hat{\pi}(t) - \hat{H}_{\bold{u}(t)}) \partial_{k} \mathcal{L}_{\bold{u}(t)}[\hat{\rho}(t)]\Big\rangle = \Big\langle  \mathcal{L}_{\bold{u}(t)}[\hat{\rho}(t)]
\partial_k\hat{H}_{\bold{u}(t)}  \Big\rangle,  \nonumber \\
\label{const} \end{eqnarray} 
where $\partial_k := \partial/ \partial u_k(t)$ is the partial derivative with respect to the $k$-th component of $\bold{u}(t)$. Finally 
using the normalization of  $\hat{\rho}(t)$ the PMP requirement {\it iii)} can be expressed as  
\begin{eqnarray} 
 \Big\langle (\hat{\pi}(t) - \hat{H}_{\bold{u}(t)}) \mathcal{L}_{\bold{u}(t)}[\hat{\rho}(t)]\Big\rangle
 =\mathcal{K}.         \label{energ1} 
 \end{eqnarray} 
It is worth remarking that  this last condition is a direct consequence of the PMP requirements {\it i)} and {\it ii)}~\cite{Kirk2012}. Still it is convenient to introduce it  because it is  an algebraic equation that allows one to neglect one degree of freedom in the differential equations  (\ref{H1}), leading
to a simplification of the calculations (cf. see next section for an application of this fact). We also notice that  condition (\ref{cons}) applies  for explicitly time independent pseudo Hamiltonians \cite{Kirk2012},
and in our case this is always true, because we relegate all possible time dependence in the controls.
 
 \section{Solution for a two-level system} 
From  Eq.~(\ref{master}) it follows that the functional~(\ref{H3i}) 
   can be expressed as 
 \begin{eqnarray}
 {\cal A}_{{u}(t)}= [2q(t) + u(t)]\frac{[e^{\beta_h u(t)}-e^{\beta_c u(t)}]}{[e^{\beta_c u(t)}+1][e^{\beta_h u(t)} +1]}\;.
 \end{eqnarray} 
 Therefore, since $u(t)\geq 0$ and $\beta_c\geq \beta_h$, we have that the kind of isotherm $S$  can experience is determined by the sign of 
 $2q(t) + u(t)$: in particular if the latter is negative then $S$  follows a
cold isotherm ($\gamma_c= \Gamma, \gamma_h=0$); if, instead, it is positive then $S$ follows a hot isotherm ($\gamma_c= 0, \gamma_h=\Gamma$).
Keeping this fact in mind 
 the PMP conditions~(\ref{const}) and (\ref{energ1}) yield the identity
  \ba \label{Kneg} 
  \mathcal{K} = - \Gamma\beta_{c,h} e^{\beta_{c,h}u(t)} \left [\frac{2q(t) + u(t)}{1+e^{\beta_{c,h}u(t)}}\right ]^2, 
  \ea
 from which it is clear that $\mathcal{K} \leq 0$ for every optimal solution.
Taking the square root of Eq.~(\ref{Kneg}) and recalling the definitions of  $\mu_{c,h}$ and $x_{c,h}(t)$ given in the main text, we have
  \begin{eqnarray} 
\label{sol2} 
2q(t) + u(t)  &=&  \frac{\mu_{c,h}}{\beta_{c,h}}\frac{[1+x_{c,h}^2(t)]}{ x_{c,h}(t)},
\end{eqnarray}
and combining this last equation with Eq. (\ref{const}) we obtain Eq.~(\ref{sol1}).
These expressions link $q(t)$ and $p(t)$ to the control field $u(t)$ when $S$ is moving along an isotherm. 
The explicit dynamical evolution finally follows from Eqs.~(\ref{dyn}) and (\ref{dyn2}), i.e. 
 \ba \label{dyna1} 
 \frac{d {p}(t)}{dt}  &=& \Gamma \Big [\frac{1}{1+x_{c,h}^2(t)} - p(t) \Big ]\;, \\
 \label{dyna222} 
 \frac{d {q}(t)}{dt}  &=& \frac{\Gamma}{2}[ 2 q(t) + u(t)]\;, 
\ea
 where in the derivation of the second expression we  used (\ref{DEFLA}) and the fact that the equilibrium state $\hat{\rho}_{eq}(t)$ associated with the dissipator~(\ref{master})
 is the Gibbs state $\hat{\eta}_{\beta_{c,h}}(t)$. 
 In particular, taking 
the time derivative of Eq.~(\ref{sol1}) and using (\ref{dyna1})   we obtain Eq.~(\ref{tdiff}) which completely determines the  time evolution of the control parameter $u(t)$ along an isotherm. Notice that the same solution can also be derived using instead  (\ref{sol2}) and (\ref{dyna222}), showing that the set of PMP conditions we are using
is indeed redundant as anticipated in the previous section. 
Before proceeding further we also remark that the above analysis assumes the gap $u(t)$ to be non negative. Yet the entire derivation can be easily extended to the case of
 negative  gaps: the only difference being in the definition of the constants 
 $\mu_{c,h}$ whose signs have to be swapped,  i.e.  $\mu_{c} = +
\sqrt{-(\beta_{c} \mathcal{K})/\Gamma}$ and  $\mu_{h} = - 
  \sqrt{-(\beta_{h} \mathcal{K})/\Gamma}$.

An implicit integration of  Eq.~(\ref{tdiff}) can be obtained  for assigned boundary conditions. 
Specifically, for a generic isothermal trajectory where $x_{c,h}(t)$ takes the values $x_0$ and $x_1$ at times $t_0$ and $t_1>t_0$ respectively,  Eq.~(\ref{tdiff}) 
imposes the following constraint 
\ba  \label{solt}   
t_1 -t_0  = 
\frac{\chi_{c,h}(x_1) - \chi_{c,h}(x_0)}{\Gamma}  \;,
\ea
with 
 $\chi_{c,h}(x):= -  (2/\mu_{c,h}) \arctan(x) + \ln [(x^2 +1)/x]$. Furthermore,  
from the definition Eq.~(\ref{heat}), we can express the heat exchanged  during such transformation as
\begin{eqnarray}  \label{finQ}  
Q_{c,h}&=& - \int_{t_0}^{t_1} dt \; u(t)  
\frac{ d{p}(t)}{dt} 
= - \int_{p_0}^{p_1} dp \;  u_{c,h}(p),  
\end{eqnarray} 
where $p_0= p(t_0)$, $p_1=p(t_1)$ and where the functions $u_{c,h}(p)$ are obtained by inverting Eq.~(\ref{sol1})  to express
the field $u$ in terms of $p$ along the isotherm, i.e. 
\begin{eqnarray} \label{invert}
u_{c,h}(p)&=&\frac{2}{\beta_{c,h}} \ln x_{c,h}(p) \;, \\ 
x_{c,h}(p) &=& \frac{\sqrt{\mu_{c,h}^2 + 4p(1-p)} -\mu_{c,h} }{2 p}\;.  \label{invert1}
\end{eqnarray} 
Equation~(\ref{finQ}) provides the geometrical interpretation of $Q_{c,h}$ as minus the area subtended by the function (\ref{invert}) as shown in Fig.~\ref{fig:2A}, furthermore 
 by direct integration can be casted in a form which is similar to~(\ref{solt}), i.e. 
 \begin{eqnarray}  \label{finQ1}  
 Q_{c,h}&=& \frac{\Xi_{c,h}(x_1) - \Xi_{c,h}(x_0)}{\beta_{c,h}}\;,
\end{eqnarray} 
 where now 
$\Xi_{c,h}(x):= -   {2\mu_{c, h}} \arctan x + [2x(x+\mu_{c, h})/(1+x^2)] \ln{x} - \ln{(1+x^2)}$.
Notice that  Eqs.~(\ref{sol1}) and (\ref{dyna1}) 
imply that for a hot isotherm $d{p}(t)/dt \geq 0$ if $x_h\in [1, 1/\mu_h]$, inside the region where $p(t)$ is defined  (i.e. $p(t) \in [0,1]$) forcing
the system to always increase the population of its excited level as explicitly indicated in  Fig.~\ref{fig:2A}. 
Accordingly, the  r.h.s.  of Eq.~(\ref{finQ}) 
 must have  $p_1\geq p_0$  making $Q_h$ negative as expected (heat must be absorbed along a hot isotherm).  
A similar reasoning holds for the cold isotherms in which the heat is always released.
Consider next the limit case where the system $S$ has enough time to fully thermalize while in contact with the reservoir, i.e. 
$\Gamma (t_1-t_0)\: \rightarrow \:\infty$. 
In this limit the only way in which the r.h.s. of Eq.~(\ref{solt}) can be arbitrarily large
if $p_0$, $p_1$ 
are fixed is by choosing $\mu_{c,h}\: \rightarrow \: 0$, i.e. $\mathcal{K} \: \rightarrow \:0$. 
With this choice it is easy to see that 
Eq.~(\ref{sol1}) reduces to the equilibrium Gibbs distribution while Eq.~(\ref{finQ}) becomes 
\begin{eqnarray} 
Q_{c,h} = \frac{[H(p_0) - H(p_1)]}{\beta_{c,h}}\label{ADI}\;,
\end{eqnarray} 
with $H(p)$ being the Shannon entropy associated to the probability $p$, that is exactly what we expect in a quasi-static process.
 
As already mentioned, in the construction of the optimal trajectory for a two-bath problem, the system $S$ may experience 
  adiabatic jumps of the control field $u(t)$ which permits  $S$  to abruptly switch  from the hot to the  cold isotherm and vice-versa. 
Such switches may occur only 
 at those special times which allow for the preservation of the continuity of  
$\hat{\rho}(t)$ and of the costate 
$\hat{\pi}(t)$  during the jump. For the two-level system case we are considering here this can be enforced by imposing the continuity $p(t)$ and $q(t)$ when
abruptly passing from the hot to the cold trajectory or vice-versa. 
Using Eqs. (\ref{sol1}) and (\ref{sol2}) we find the two conditions
\begin{eqnarray}  \label{ad1.0} 
\frac{1-\mu_{c} x_{c}(p)}{1+x_{c}^2(p)}&=&\frac{1- \mu_{h} x_{h}(p)}{1+x_{h}^2(p)}, \\ 
u_c(p) - \frac{\mu_c[1+x_c^2(p)]}{\beta_c x_c(p)}&=&u_h(p) -\frac{\mu_h[1+x_h^2(p)]}{\beta_hx_h(p)}. \nonumber \\\label{ad2} 
   \end{eqnarray} 
Given the functional dependence of  Eqs.~(\ref{invert}) and (\ref{invert1}) it follows that (\ref{ad1.0}) is always satisfied for all values of $p$. Equation~(\ref{ad2}) instead selects
the values of $p$ which fulfill the 
 following condition  
 \begin{eqnarray}\label{ad1}
 f({p}; \mathcal{K})=0, 
 \end{eqnarray}  
 with 
\ba \label{adiabfin} 
f({p}; \mathcal{K})&:=& \frac{(\Delta_c-\mu_c)}{(\Delta_h-\mu_h)} - \frac{(\Delta_h-\mu_h)}{(\Delta_c-\mu_c)}  
+2\sqrt{\beta_c\beta_h}
\nonumber \\
&\times &
\biggl [\frac{1}{\beta_c}\ln\frac{(\Delta_c -\mu_c)}{2p}-  \frac{1}{\beta_h}\ln\frac{(\Delta_h-\mu_h)}{2p}   \biggr ],  
\ea
and  $\Delta_{c, h}({p}) := \sqrt{\mu^2_{c, h} +4{p}(1-{p})}$.
It is easy to show that  for each value of ${p}$ and for each choice of $\beta_c$ and $\beta_h$, $f({p}; \mathcal{K})$ is monotonically decreasing in $\mathcal{K}$.
Thus for the implicit function theorem the zero contour of $f({p},\mathcal{K})$ is the graph of a function of ${p}$ which we plot
in Fig.~\ref{fig:2B}a. 
One notices that  for each value of $\mathcal{K}>\mathcal{K}^*$ which depends upon $\beta_{c,h}$ and $\Gamma$,  there are always two values of ${p}$ nullifying $f({p}; \mathcal{K})$ and identifying the trajectory points where the system can switch from
one isothermal regime to the other, while no transitions are allowed for values of $\mathcal{K}<\mathcal{K}^*$. 
By close inspection of the r.h.s. of Eq.~(\ref{adiabfin}) one may notice that ${\cal K}$ appears in it always through the ratio ${\cal K}/\Gamma$.
Furthermore  a simple dimensional analysis of the same equation can be used to verify that the quantity $\beta_c f(p,{\cal K})$ is  a function of the ratio $\beta_h/\beta_c$. Equation (\ref{kstar}) follows from these two observations.

\begin{figure*}[t!]
\includegraphics[width=0.95\textwidth]{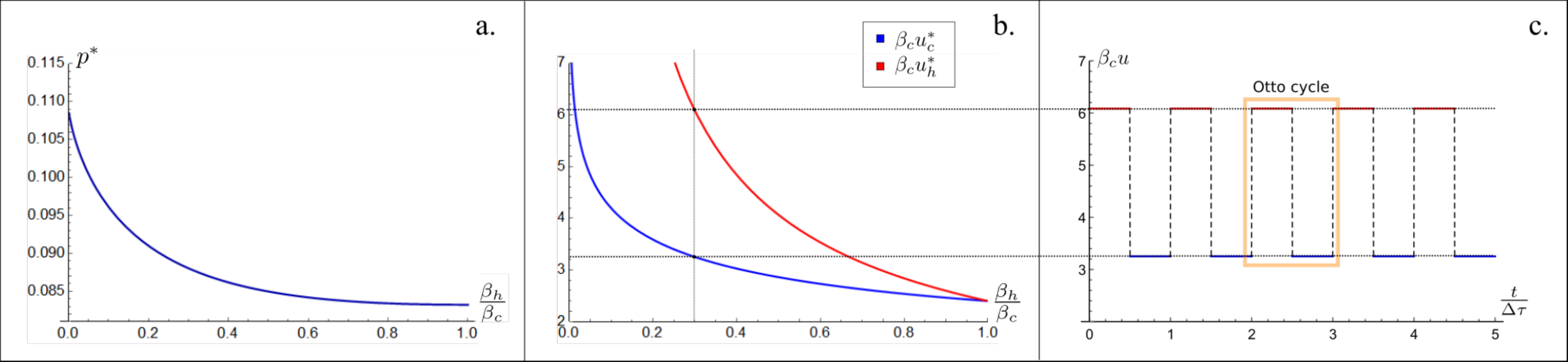}
 \caption{
 {\bf a.} Optimal excitation probability $p^*=p_{ad, 1}({\cal K}^*)=p_{ad, 2}({\cal K}^*)$ as a function of the temperature ratio.
{\bf b.} The two extrema of the quenches characterizing the maximum power heat engine $u_c^*$ and $u_h^*$,
measured in units of $\beta_c^{-1}$ as a function of $\beta_h/\beta_c$. The case of $\beta_h / \beta_c=0.3$ is singled out.
{\bf c.} The optimal driving protocol for the parameter $u^*(t)$, measured in units of $\beta_c^{-1}$.
The driving is composed by a succession of infinitesimal Otto cycles, where the extrema $u_c^*$ and $u_h^*$ can be found directly from the left panel of the figure.
The value $\Delta \tau \ll \tau$ is an arbitrarily small unit of time expressing the duration of each degenerate cycle.}
\label{fig:4}
\end{figure*}

\section{Dependency of optimal strategies on $\mathcal K$} 
Differentiating both terms of Eq.~(\ref{solt})  with respect to ${\cal K}$ we get
\begin{eqnarray}  \label{dqdk} 
\pd{}{\cal K}(t_1-t_0) &=& - \left(\frac{\beta_{c,h}}{2\mu_{c,h} \Gamma} \right)
\pd{}{\mu_{c,h}}(t_1-t_0) \nonumber \\
&=& \frac{[ \arctan (x_1)-\arctan (x_0)] }{ \Gamma {\cal K}\mu_{c,h}}.
\end{eqnarray} 
Now for a cold  isotherm we have that $p(t)$ is a decreasing function of time, while $u(t)$, and hence $x(t)$, are
increasing  (see Eq.~(\ref{solt})).
Accordingly in the above equation we have $x_1 \geq x_0$. Exploiting  the fact that  $\arctan[x]$ is an increasing function 
of its argument and  the fact that $-\mu_c$ is a positive quantity, we can conclude that for a cold isotherm the time interval $t_1-t_0$ it takes the system to move from
one point of the trajectory to the other, is an increasing function of ${\cal K}$. Exactly the same conclusion can be inferred for the hot isotherm (here
the fact that $x_1 \leq x_0$ is compensated by the negativity of $-\mu_h$). Using this observation we can then conclude that 
 the quantities $\tau_c(\mathcal{K})$, $\tau_h(\mathcal{K})$
and $\tau_{cycle}(\mathcal K)$ appearing in Eq.~(\ref{dect}) are 
 increasing functions of $\mathcal{K}$.
Similarly, for the quantities appearing on the r.h.s. of Eq.~(\ref{decQ}) we simply use Eq.~(\ref{finQ1}) 
to observe that 
\ba \label{qkt}
 \pd{Q_{c,h}}{\cal K} ={\mathcal{K}}\pd{}{\cal K}(t_1-t_0),
\ea
and we then conclude that $Q_c(\mathcal{K})$, $Q_h(\mathcal{K})$
and $Q_{cycle}(\mathcal K)$ are decreasing functions of $\mathcal{K}$.

\section{Evaluation of $\mathcal K^*$ and $p^*$} 
For a two-level system which can be put into contact with either of two heat baths of fixed temperatures $\beta_c$ and $\beta_h$, the maximum power (or minimum dissipation) control strategy is completely determined by two parameters: the optimal rate $\mathcal K^*$ and the optimal excitation probability $p^*$. Indeed, according to the results presented in the main text, the maximum power $-\mathcal K^*$ is achieved by a sequence if infinitesimal Otto-cycles performed around the non-equilibrium working point $p^*$. In this regime the energy gap is periodically switched between the two finite values $u_c(\mathcal K^*,p^* )$ and $u_h(\mathcal K^*,p^*)$ (given by \eqref{invert}, \eqref{invert1}), each quench being followed by an incomplete thermalization with the cold and hot bath respectively. 
From the previous analysis we know that the desired optimal parameters $\mathcal K^*$ and $p^*$ are solutions of the following  minimization problem: $\min_{~\mathcal K \in A ~;~ p}  \mathcal K$
subject to condition \eqref{ad1}, determining the accessible region of the control problem and represented in Fig. \ref{fig:2B}a.
This minimization problem is efficiently determined from the following system of algebraic equations 
\begin{equation} \label{sys}
\left\{
\begin{array}{ccc}
\quad f(\mathcal K,p) &=& 0, \\
\frac{\partial}{\partial p} f(\mathcal K,p) &=& 0 ,
\end{array}
\right.
\end{equation}
where, after some calculations, the second equation can be explicilty written as:
\begin{equation} \label{final}
\frac{[1 + x^2_c(\mathcal K,p )]}{\mu_c(\mathcal K,p ) x_c(\mathcal K,p )} = - \frac{[1 + x^2_h(\mathcal K,p )]}{\mu_h(\mathcal K,p ) x_h(\mathcal K,p) }.
\end{equation}
Expressing $\mathcal{K}$ in units of $\Gamma/\beta_c$, it is easy to check that the solution of~(\ref{sys}) depends only on the temperature ratio $\beta_h/\beta_c$,
so that it could be casted as in Eq.~(\ref{kstar}).
The strictly decreasing nature of $g(z)$ suggests that optimal
thermal engines are particularly 
performant in the limit $z\rightarrow 0$.
Indeed in this regime a machine
working with fixed temperature gradient $\delta=  \beta_h^{-1} -  \beta_c^{-1} $ produces the maximum
power output.
This is verified by direct substitution of $y:=(\delta \beta_c)^{-1}$ and $z= y/(1+y)$ 
in Eq.~(\ref{kstar}),  and noticing  that the function $y g[y/(1+y)]$ is also
 monotonically decreasing in $y$.
This limit can be also treated analytically, solving the system (\ref{sys}) for infinitesimal
values of $\beta_h/\beta_c$.
Using the asymptotic behaviour $g(z) \approx \theta/z$, replacing Eq.~(\ref{kstar}) 
into Eq.~(\ref{final}), and 
considering only the leading order coefficients in the expansion for small $z$ of we obtain
\ba \label{pteta} p^*(z\rightarrow 0) = \frac{2 \theta}{1+4\theta}. \ea
To complete the solution of  (\ref{sys}) we apply the same expansion procedure to Eq.~(\ref{ad1})
which, combined with (\ref{pteta}), gives 
\ba  \label{eqteta}  4\theta e^{4\theta} = e^{-1},  \ea
whose solution can be expressed in terms of the so called Lambert function $W$ \cite{Corless1996}:
\ba  \label{theta}
\theta= \frac{W(e^{-1})}{4} \approx 0.06961. 
\ea
Equation (\ref{theta})  confirms our numerical result for $\theta$, and makes it possible to estimate
$p^*$ through Eq.~(\ref{pteta}), yielding  $p^*(z\rightarrow 0)= 0.10848$.
This is in agreement with the numerical 
plot for the optimal excitation probability $p^*$, shown in Fig. \ref{fig:4}a for all temperature ratios.
Once the values of the quantities $\mathcal{K}=\mathcal{K}^*$ and $p=p^*$ have been determined, one can also find the extremal
values $u^*_{c,h}$ of the maximum power Otto cycle by simply inverting the functional dependence that links $p$ and $u$ in Eq.~(17) (see Fig. \ref{fig:4}b).
Finally, the optimal control protocol for the parameter $u^*(t)$ can be found from the optimal values of $\mathcal{K}^*, p=p^*,
u^*_{c,h}$ through Eq.~(34) (see Fig. \ref{fig:4}c).
Noticeably the time durations of the two isochores are equal, as
could be directly verified from eqs. (\ref{dyna1}) and (\ref{ad1}).


\begin{thebibliography}{1}
\bibitem{NOTA2} With the word {\it isotherm} we refer to transformations in which the system is in contact with only one bath at a time, even if the evolution is not quasi-static.
\bibitem{NOTA} Notice that since $\langle \mathcal{L}_{\bold{u}(t)} \hat{\rho}(t)\rangle= \langle d\hat{\rho}(t)/dt\rangle=0$,
 there is some arbitrariness in the definition of ${\hat{\pi}}(t)$ which can be shifted by adding an arbitrary time-dependent real function. 
For simplicity we fix this freedom by choosing the gauge $\langle{\hat{\pi}}(t)\rangle=0$ for all $t\in[0,\tau]$.
\end{thebibliography}
\end{document}